\documentclass[eqsecnum,aps,twocolumn,epsf,showpacs]{revtex4} % PH. REV.
\usepackage{graphicx}
\begin{document}
   
%\twocolumn[\hsize\textwidth\columnwidth\hsize\csname@twocolumnfalse\endcsname
 
\title{Evolution
of a collapsing and exploding Bose-Einstein condensate in different trap
symmetries}
 
\author{Sadhan K. Adhikari}
\affiliation{Instituto de F\'{\i}sica Te\'orica, Universidade Estadual
Paulista, \\ 01.405-900 S\~ao Paulo, S\~ao Paulo, Brazil\\}

\date{\today}
 
%\maketitle
 
\begin{abstract}
 
Based on the time-dependent Gross-Pitaevskii equation we study the
evolution of a collapsing and exploding Bose-Einstein condensate in
different trap symmetries to see the effect of confinement on collapse and
subsequent explosion, which can be verified in future experiments.
 We make prediction for the evolution of the shape of the condensate and
the number of atoms in it for different trap symmetries (cigar to pancake)
as well as in the presence of an optical lattice potential.  We also make
prediction for the jet formation in different cases when the collapse is
suddenly terminated by changing the scattering length to zero via a
Feshbach resonance.

\end{abstract}

\maketitle

\section{Introduction}

Since
the detection and study of Bose-Einstein condensates 
(BECs) of $^7$Li atoms with attractive interaction \cite{ex2}, 
such  condensates
have been used in the study of solitons \cite{sol}  and collapse
\cite{don}.
In general an
attractive condensate with number of atoms $N$
larger than a critical value $N_{\mbox{cr}}$
is not dynamically
stable  \cite{ex2}. However, if such a strongly attractive 
condensate is ``prepared"  or somehow made to exist it experiences a
dramatic collapse and explodes 
emitting atoms. The first demonstration of such a collapse was made with a 
$^7$Li condensate by slowly increasing the number of atoms in it from an
external source, while the BEC showed a sequence of collapse with the
number of atoms $N$ oscillating around $N_{\mbox{cr}}$.  Such a collapse
is driven by a stocastic process.

A dynamical study of a much stronger and violent  collapse   
has been performed by Donley {\it et al.}
\cite{don} on an attractive  $^{85}$Rb BEC \cite{ex3}
in an axially
symmetric trap, where they
manipulated the inter-atomic interaction by changing the external magnetic
field exploiting  a nearby Feshbach resonance \cite{fs}. 
In the vicinity
of a Feshbach resonance the atomic scattering length $a$ can be varied
over a huge range by adjusting the external magnetic field.  
Consequently,
they  changed the sign of the scattering length, thus
transforming a
repulsive condensate of $^{85}$Rb atoms 
into an attractive one which naturally evolves into a collapsing and
exploding condensate. 
Donley {\it et al.}
 provided a quantitative estimate of the explosion of this BEC
by measuring different properties of the exploding condensate.

It has been realized that many features of the experiment by Donley {\it
et al.} \cite{don} on the collapsing condensate 
can be described 
\cite{th1,th2,th2a,th3,th3a,th4,th5,th7,th9,th10,th11,sk5,sk6} 
by the mean-field Gross-Pitaevskii
(GP) equation \cite{11}.
To account for the loss of atoms from the
strongly attractive collapsing condensate an absorptive
nonlinear three-body recombination term is included in the GP equation
\cite{th1}.
However, we are fully aware that there are features of this experiment
which are expected to be beyond mean-field description. Among these are
the distribution of number and energy of emitted high-energy ($\sim
10^{-7}$ Kelvin) uncondensed burst atoms reported in the experiment.
Although there have been some attempts \cite{th3,th3a,th4} to describe the
burst atoms using the mean-field GP equation, now there seems to be a
consensus that they cannot be described adequately and satisfactorily
using a mean-field approach \cite{th7,th9,th10}.  Also, the 
GP equation does not successfully predict the ``time to collapse" (or the 
time lag to start the collapse after changing the sign of the scattering
length) 
in all
cases investigated in the experiment, as has been pointed out in Refs. 
\cite{th5,th11}. 

The GP equation is supposed to deal with the zero- or very low-energy
condensed phase of atoms and has been  used to predict the
time to collapse, evolution of the collapsing condensate including the
very low-energy ($\sim$ nano Kelvin) jet
formation \cite{don} when the collapse is suddenly stopped before
completion by jumping the scattering length to $a_{\mbox{quench}}=0$
 (noninteracting atoms)
or positive (repulsive atoms) values. 
The jet atoms are slowly formed in the radial
direction
when
the collapse is stopped  in this fashion. 
In the experiment usually
$a_{\mbox{quench}}=0$. It is emphasized that unlike the
emitted 
uncondensed ``hotter"
missing
and burst atoms reported in the experiment \cite{don} the jet atoms form a
part of the surviving ``colder" condensate and hence should be describable
by
the mean-field GP equation. 
Saito {\it et al.}  \cite{th3}, Bao {\it et al.} 
\cite{th10} and this author \cite{sk5}
 presented a mean-field description of jet formation and
Calzetta
{\it et al.}       \cite{th9}
treated jet
formation \textit{exclusively} as a quantum effect.
More recently, the present author has used a set of coupled
mean-field-hydrodynamic 
equations \cite{sk6} to describe the
essentials of the collapse dynamics of a mixture of a boson and fermion
condensates \cite{bf}.

In this paper we extend the study of the evolution of the collapsing and
exploding
condensate in different symmetries to see the effect of confinement on
collapse and subsequent explosion.
Future experiments may verify these predictions and thus
provide a more stringent test for the mean-field GP equation. 
The experiment of  Donley {\it et al.} was
performed for an axially-symmetric   cigar-shaped BEC. In the present
analysis we extend
our study to a spherical as well as an  axially-symmetric  pancake-shaped
BEC.  

Lately, the periodic optical-lattice potential has played an essential
role in many theoretical and experimental studies of  Bose-Einstein
condensation, e. g., in the study of Josephson oscillation \cite{j} and
its disruption \cite{jd},
interference
of matter-wave \cite{i}, BEC dynamics on periodic trap \cite{d}, etc. 
The optical-lattice confinement creates a BEC in an entirely different
shape and trapping condition form a conventional harmonic oscillator
trapping.  Consequently, one could have a collapse of a different nature
in the presence of an optical-lattice potential. We shall see in our study  
that under certain conditions of trap symmetry,  in addition to the usual
global collapse to the centre, in the presence of the optical-lattice  
potential
one could have independent local collapse
of pieces of
the condensate  to local centres. 
In
view of
this we
study the dynamics of a collapsing and exploding BEC of different
symmetries 
prepared on a
periodic optical-lattice potential.  We study the evolution of
the shape  and size of the condensate as well as the jet formation upon
stopping the collapse by making the BEC repulsive or noninteracting.

In Sec. II we present our  mean-field model. In Sec. III we present
our results that we compare with the
experiment and other numerical studies. In Sec. III  we also present a
physical
discussion of our findings 
and some concluding remarks  are given in Sec. IV.
 
\section{Nonlinear Gross-Pitaevskii Equation}

The time-dependent Bose-Einstein condensate wave
function $\Psi({\bf r};\tau)$ at position ${\bf r}$ and time $\tau $
allowing
for atomic loss
may
be described by the following  mean-field nonlinear GP equation
\cite{11}
\begin{eqnarray}\label{a}& \biggr[& - i\hbar\frac{\partial
}{\partial \tau}
-\frac{\hbar^2\nabla^2   }{2m}
+ V({\bf r})
+ gN|\Psi({\bf
r};\tau)|^2-  \frac{i\hbar}{2}
\nonumber \\
& \times & (K_2N|\Psi({\bf r};\tau) |^2
+K_3N^2|\Psi({\bf r};\tau) |^4)
 \biggr]\Psi({\bf r};\tau)=0. \nonumber \\
\end{eqnarray}
Here $m$
is
the mass and  $N$ the number of atoms in the
condensate,
 $g=4\pi \hbar^2 a/m $ the strength of inter-atomic interaction, with
$a$ the atomic scattering length. 
The terms $K_2$ and $K_3$ 
denote two-body
dipolar and three-body recombination loss-rate coefficients, respectively
and include the Bose statistical factors 
$1/2!$ and $1/3!$ needed to describe the condensate.
The trap potential with cylindrical symmetry may be written as  $  V({\bf
r}) =\frac{1}{2}m \omega ^2(\rho^2+\nu^2 z^2)+ V_{\mbox{op}}$ where
 $\omega$ is the angular frequency
in the radial direction $r$ and
$\nu \omega$ that in  the
axial direction $z$ of the harmonic trap. The cigar-shaped condensate
corresponds to $\nu <1$ and pancake-shaped condensate corresponds to 
$\nu >1$.
The 
periodic optical-lattice potential in the axial $z$
direction
created by a standing-wave laser field of
wave length $\lambda$ is given by $V_{\mbox{op}}=\kappa E_R \cos ^2 (k_L
z)$ with $E_R=\hbar^2 k_L^2/(2m), k_L=2\pi/\lambda$ and $\kappa$ the
strength. 
The normalization condition of the wave
function is
$ \int d{\bf r} |\Psi({\bf r};\tau)|^2 = 1. $
Here we
simulate the atom  loss via
the most important quintic three-body term  $K_3$ \cite{th1,th2,th2a,th3}.
The contribution of the cubic  two-body  loss term  $K_2$
\cite{k3} is
expected to be negligible \cite{th1,th3} compared to the  three-body term
in
the present problem of the  collapsed condensate with large density
and will not be considered here.

In the absence of angular
momentum the wave function has the form $\Psi({\bf
r};\tau)=\psi(\rho,z;\tau).$
Now  transforming to
dimensionless variables
defined by $x =\sqrt 2 \rho/l$,  $y=\sqrt 2 z/l$,   $t=\tau \omega, $
$l\equiv \sqrt {\hbar/(m\omega)}$,
and
\begin{equation}\label{wf}
\phi(x,y;t)\equiv
\frac{ \varphi(x,y;t)}{x} =  \sqrt{\frac{l^3}{\sqrt 8}}\psi(\rho,z;\tau),
\end{equation}
we get
\begin{eqnarray}\label{d1}
\biggr[&-&i\frac{\partial
}{\partial t} -\frac{\partial^2}{\partial
x^2}+\frac{1}{x}\frac{\partial}{\partial x} -\frac{\partial^2}{\partial
y^2}
+\frac{1}{4}\left(x^2+\nu^2 y^2-\frac{4}{x^2}\right)\nonumber \\
& +&\kappa \frac{4\pi^2}{\lambda_0 ^2}
\cos^2( \frac{2\pi}{\lambda_0}y   )
+ 8 \sqrt 2 \pi   n\left|\frac {\varphi({x,y};t)}{x}\right|^2 \nonumber \\    
&-& 
i\zeta n^2\left|\frac {\varphi({x,y};t)}{x}\right|^4
 \biggr]\varphi({ x,y};t)=0,
\end{eqnarray}
where
$ n =   N a /l$, $\lambda_0=\sqrt
2\lambda/l$
and $\zeta =4K_3/(a^2l^4\omega).$
The normalization condition  of the wave
function becomes
\begin{equation}\label{5} {\cal N}_{\mbox{norm}}\equiv {2\pi} \int_0
^\infty
dx \int _{-\infty}^\infty dy|\varphi(x,y;t)|
^2 x^{-1}.  \end{equation}
For $\zeta =K_3=0,$  ${\cal N}_{\mbox{norm}}=1$, however, in the presence
of
loss
$K_3 > 0$, ${\cal N}_{\mbox{norm}}
< 1.$ The number of remaining atoms $N$
in the condensate is given by $ N=N_0
{\cal N}_{\mbox{norm}}$, where $N_0$ is the initial number of atoms.

In this study the term $K_3$ or $\zeta =4K_3(a^2l^4\omega)$ 
will be used for a description of atom loss
in the case of attractive interaction. The choice of $K_{3}$ has a huge
effect on some experimental  observables and the
fact that it is experimentally not precisely determined is a
problem for existing theory on the experiment.
As in our previous study
\cite{sk5} we employ $\zeta  =2$ and  $K_3\sim a^2$  throughout this
study.
It was found \cite{sk5} that this value of $\zeta  (=2)$ reproduced  the
time
evolution of the
condensate in the experiment of Donley {\it et al.} \cite{don} 
satisfactorily for a wide range of variation of initial number of atoms
and scattering lengths
\cite{th2}.
The present value  $\zeta =2$  with $K_3=\zeta  a^2l^4\omega/4$  leads to
\cite{th2,th2a}
$K_3\simeq 8\times 10^{-25}$ cm$^6$/s at $a=-340a_0$ and  
$K_3\simeq 6\times 10^{-27}$ cm$^6$/s at $a=-30a_0$, where $a_0$ is the
Bohr radius.  
The experimental value  of loss rate is \cite{k3}
$K_3\simeq 7\times 10^{-25}$ cm$^6$/s at $a=-340a_0$ which is
very close to the present choice. 
Of the theoretical studies, the $K_3$ values used by Santos 
{\it et al.}
\cite{th4}  ($K_3\simeq 7\times 10^{-25}$ cm$^6$/s at $a=-340a_0$),
Savage {\it et al.} \cite{th5} ($K_3\simeq 19\times 10^{-27}$ cm$^6$/s
at $a=-30a_0$),
Bao 
{\it et
al.} \cite{th10} 
($K_3\simeq 6.75\times 10^{-27}$ cm$^6$/s at $a=-30a_0$) and the present
author \cite{th2} are
consistent with each other 
and describes well the decay of the collapsing condensate.

\section{Numerical Result}

\begin{figure}[!ht]
 
\begin{center}
\includegraphics[width=0.95\linewidth]{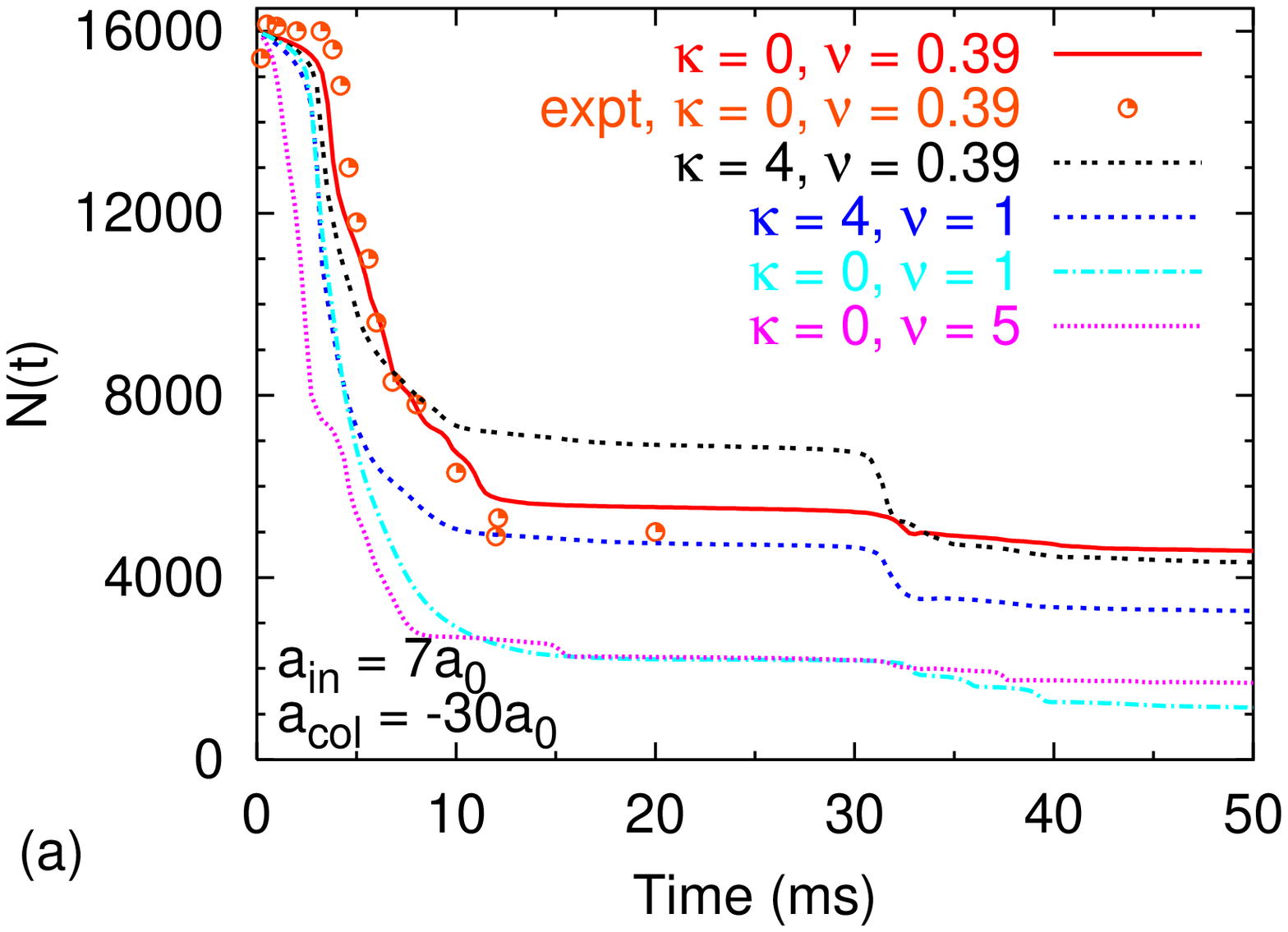}
\includegraphics[width=0.95\linewidth]{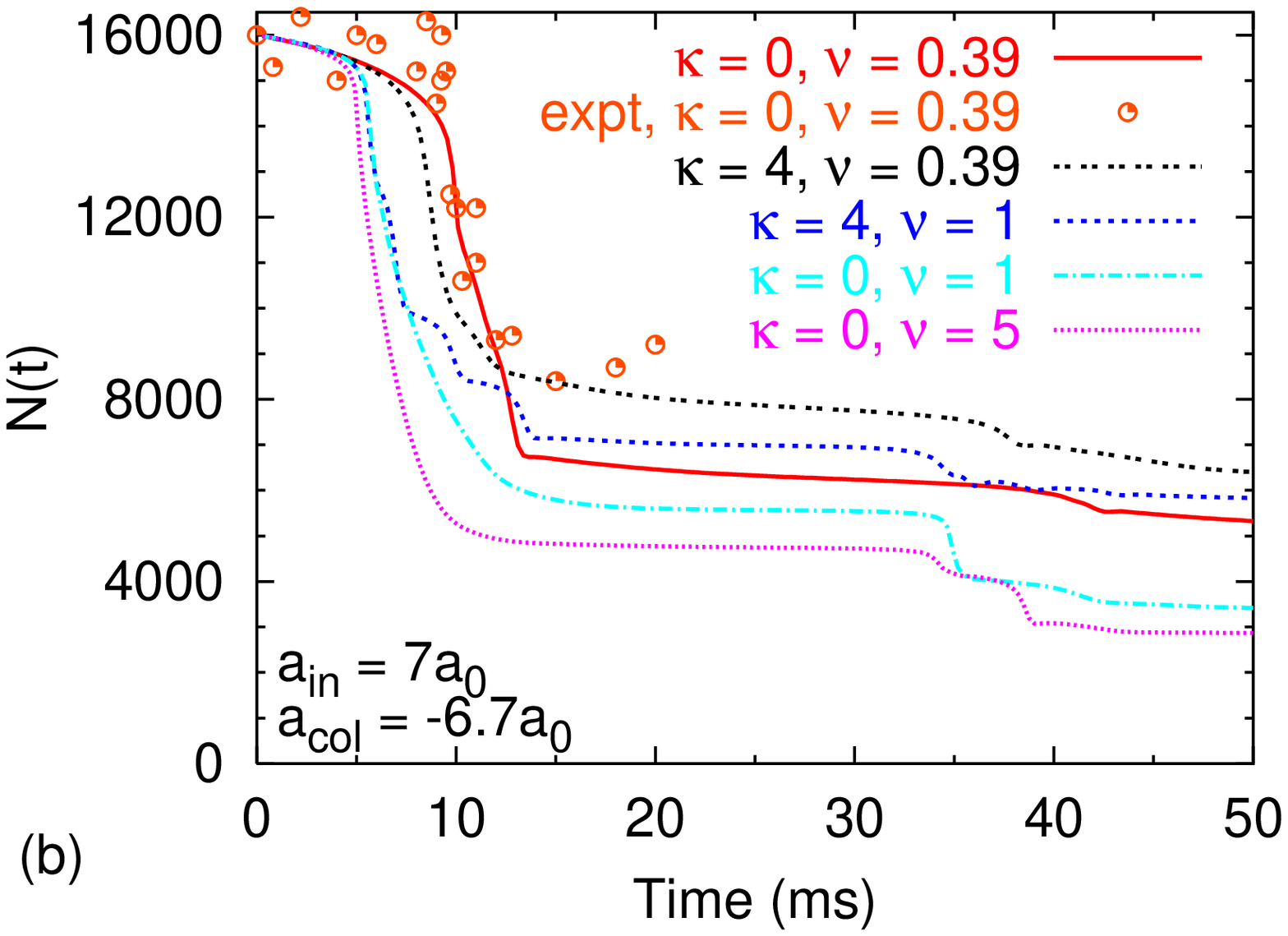}
\end{center}

\caption{(Color online) Number of remaining atoms $N(t)$ in the condensate
of 16000
$^{85}$Rb
atoms after ramping the scattering length from $a_{\mbox{in}}=7a_0$ to 
(a)  $a_{\mbox{col}}=-30a_0$ and (b)  $a_{\mbox{col}}=-6.7a_0$ as a
function of evolution time in milliseconds. The unpublished and unanalyzed 
experimental 
points of Donley {\it et al.} \cite{don}  for  $a_{\mbox{col}}=-6.7a_0$
are
taken from Bao {\it et al.}   \cite{th10}.
The curves are labeled by their respective optical lattice strength
$\kappa$ and axial trap parameter $\nu$.}

\end{figure}
 
We solve the GP equation (\ref{d1}) numerically using a time-iteration
method based on the Crank-Nicholson discretization scheme elaborated in
\cite{sk1}.  We discretize the GP equation using time step $\Delta=0.001$
and space step $0.1$ for both $x$ and $y$ spanning $x$ from 0 to 15 and
$y$ from $-30$ to 30. This domain of space was sufficient to encompass the
whole condensate wave function in this study.

\begin{figure}[!ht]
 
\begin{center}
\includegraphics[width=0.95\linewidth]{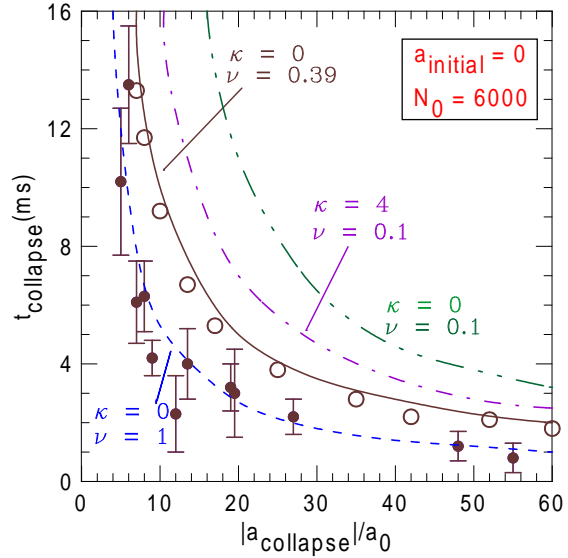}
\end{center}

\caption{(Color online) The time to collapse $t_{\mbox{collapse}}$ vs  
$|a_{\mbox{collapse}}|/a_0 $ for $a_{\mbox{initial}}=0$, $N_0=6000$. 
Solid circle with error bar: experiment \cite{don} with $\nu = 0.39$ and
$\kappa=0$; Open circle: mean-field model of \cite{th3} with $\nu = 0.39$
and
$\kappa=0$; full line: present result with  $\nu=0.39$ and
$\kappa=0$;
dashed line: present result with  $\nu = 1$ and
$\kappa=0$; dashed-dotted line: present result with  $\nu = 0.1$ and
$\kappa=4$; and dashed-doubled-dotted  line: present result with  $\nu =
0.1$ and
$\kappa=0$. }

\end{figure}

First, the numerical simulation is performed with the actual parameters of
the
experiment by Donley {\it et al.} \cite{don}, e. g., the initial number of
atoms, scattering lengths, etc. Throughout this investigation we take 
the harmonic oscillator length 
$l=2607$ nm \- and one unit of time $t$ = 0.009 095 s \cite{th2}
consistent with the experiment of Donley {\it et al.} \cite{don}. When
we include an optical-lattice potential, the
optical-lattice strength $\kappa$ is taken to be 4, and the reduced wave
length $\lambda_0$ is taken to be 1 throughout this study.  These
optical-lattice parameters are consistent with the experiment by
Cataliotti {\it et al.} \cite{j,jd}. The numerical
simulation using Eq. (\ref{d1}) with a nonzero $\zeta  (=2)$ immediately
yields
the remaining number of atoms in the condensate after the jump in
scattering length. 

\subsection{Evolution of the Number of Atoms in the Condensate}

In the experiment the initial scattering length 
$a_{\mbox{in}} (>0)$ of a repulsive condensate is suddenly jumped   to
$a_{\mbox{col}} (<0)$  to start the collapse. 
The remaining number $N(t)$ of atoms vs. time for an initial
number of atoms $N_0=16000$ and an initial scattering length
$a_{\mbox{in}}=7a_0$ are shown in Figs. 1 (a) and (b) for final
scattering lengths after collapse $a_{\mbox{col}}=-30a_0$ and $-6.7a_0$,
respectively. In both cases the experimental data for $\kappa =0$ and $\nu
=0.39$ (cigar-shaped condensate) are in agreement with the theoretical
simulation without any adjustable parameter.  For
$a_{\mbox{col}}=-6.7a_0$, 
the unpublished experimental data of \cite{don} as shown in 
Fig. 1 (b) 
are as quoted in  Bao {\it
et
al} \cite{th10}. These data are not fully analyzed and for large time 
are expected to be bigger than the actual number of atoms. This is due to
the difficulty in separating the remnant condensate from the oscillating
atom cloud surrounding it \cite{don}.  In addition, in Figs. 1 we plot
the
results for $\kappa =4$ and $\nu =0.39$ (cigar-shaped condensate with
optical-lattice potential);  $\kappa =4$ and $\nu =1$ (spherical
condensate with optical-lattice potential); $\kappa =0$ and $\nu =1$
(spherical condensate); and $\kappa =0$ and $\nu =5$ (pancake-shaped
condensate).

As the repulsive condensate is quickly turned attractive at $t=0$, via a
Feshbach resonance, the condensate starts to collapse and once the central
density increases sufficiently it loses a significant portion of atoms in
an explosive fashion via three-body recombination to form a remnant
condensate in about 15 ms as can be seen in Figs. 1. After explosion 
the number of atoms
in the remnant continues to be 
 much larger than the critical number of atoms
$N_{\mbox{cr}}$ and it keeps on losing atoms at a much slower rate without
undergoing violent explosion. However, in some cases the remnant undergoes
a smaller secondary explosion while it loses a reasonable fraction of
atoms
in
a small interval of time. This happens when the number of atoms in the
remnant is much larger than $N_{\mbox{cr}}$ so as to initiate a secondary
collapse and explosion. Prominent secondary
explosions in the presence of optical-lattice potential are found in
different cases in 
Figs.  1  for $40>t>30.$

\subsection{Time to Collapse}

Another important aspect of collapse is the ``time to collapse" or the
time to initiate the collapse and explosion  $t_{\mbox{collapse}}$
after the repulsive
condensate is suddenly
made attractive at $t=0$. Collapse is characterized by a sudden rapid
emission of atoms from the condensate. From Figs. 1 we find that the
time to collapse is the shortest for a pancake-shaped symmetry
($\nu>1$) and is the longest for a cigar-shaped symmetry ($\nu < 1).$ The
inclusion of an optical-lattice potential has no effect on the time to
collapse for a spherical or pancake-shaped symmetry. However, its
inclusion reduces the time to collapse for a cigar-shaped
symmetry. These features of time to collapse are illustrated  in Fig. 2
where we
plot $t_{\mbox{collapse}}$ vs.  $|a_{\mbox{collapse}}|/a_0$
 of the 
collapse of a condensate of 6000 atoms originally in a
noninteracting state with scattering length $a_{\mbox{initial}}=0$. 
Then suddenly its scattering length is changed to a negative
(attractive) value $a_{\mbox{collapse}}$ and its  $t_{\mbox{collapse}}$ is
obtained. Donley {\it et al.} experimentally measured
$t_{\mbox{collapse}}$ 
in this case for $\nu = 0.39$ and $\kappa=0$ and here we provide the same
for other
values of trap symmetry $\nu$ and also in the presence of a optical
lattice potential with $\kappa =4$. It should be recalled that  the
prediction of the GP equation by this author and others
\cite{th3,th5,th11}
does not very well describe the
experimental results of  Donley {\it et al.} for the time to collapse.
The inclusion of the optical-lattice
potential has reduced the time to collapse in a cigar shaped condensate
($\nu =0.1$).

The above features  of time to collapse could be understood on a physical
ground. In a cigar-shaped condensate the average distance among the atoms
is larger than that in a pancake-shaped condensate of same volume. Hence,
due to atomic attraction a cigar-shaped condensate has to contract during
a larger interval of time than a pancake-shaped condensate before the
central density increases sufficiently to start an explosion. This
justifies a larger time to collapse for a cigar-shaped condensate.  
In the presence of an optical-lattice potential for 
cigar-shaped symmetry the optical-lattice divides the condensate
in a large number of pieces. What predominates in the collpase of such a
condensate 
is the collapse of an individual piece to a local  center rather
than to the global center of the condensate via tunneling. This is a
quicker
process than the collapse of the whole condensate to the
global center. This is
why the time to collpase is shorter for a cigar-shaped condensate in an
optical-lattice trap than a cigar-shaped condensate in a harmonic trap
alone.  In a pancake-shaped symmetry the number of optical-lattice sites
inside the condensate is small. In this case a separation of the
condensate in a smaller number of  pieces
does not
aid in the collapse, as the different slices of the condensate has to
collapse essentially 
towards the center of the condensate before the explosion starts.
Hence the optical-lattice potential has   almost  no effect on the time
to collapse in the pancake-shaped or spherical symmetry.

\begin{figure}%[!ht]
 
\begin{center}
\includegraphics[width=0.7\linewidth]{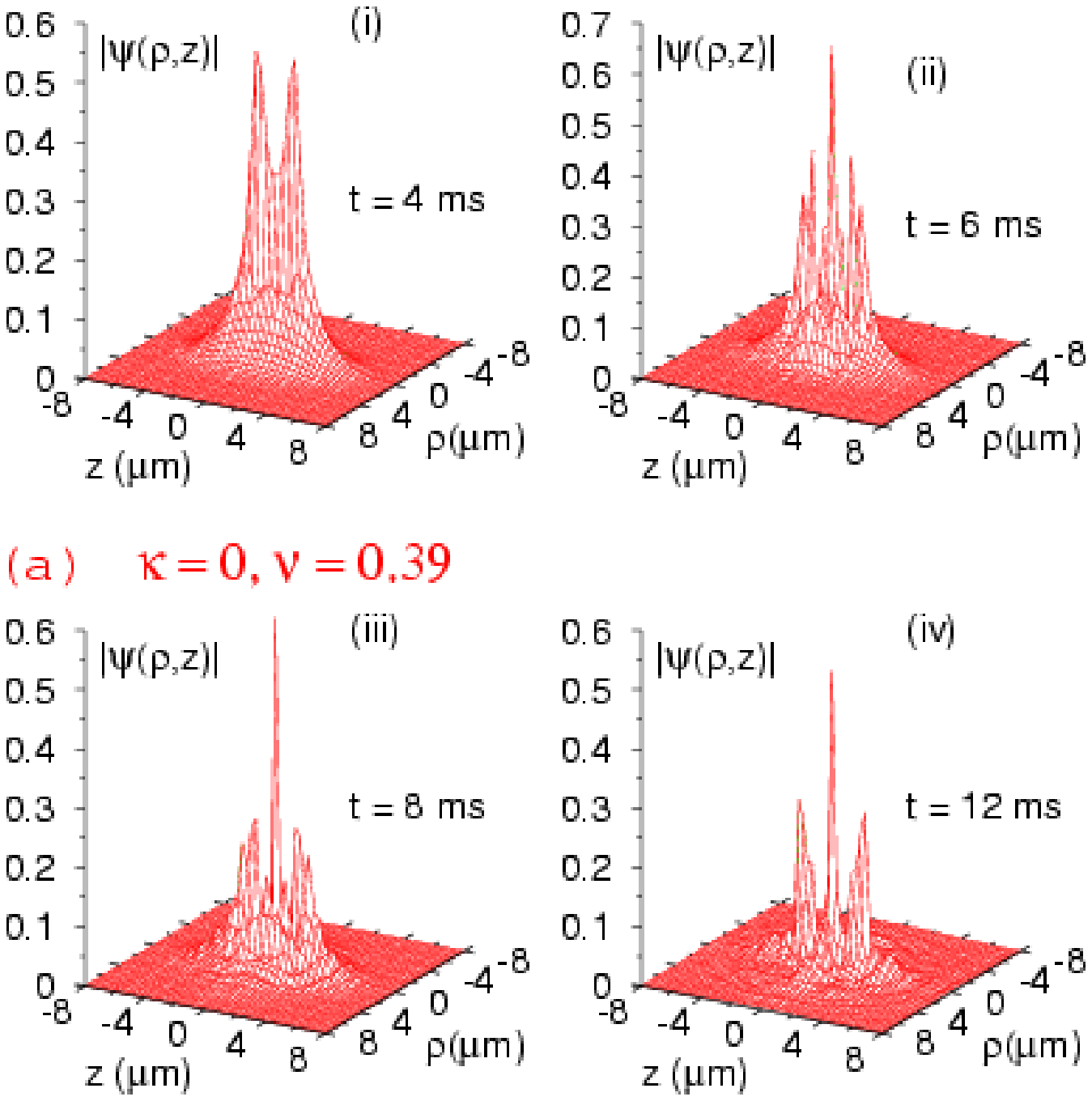}
\includegraphics[width=0.7\linewidth]{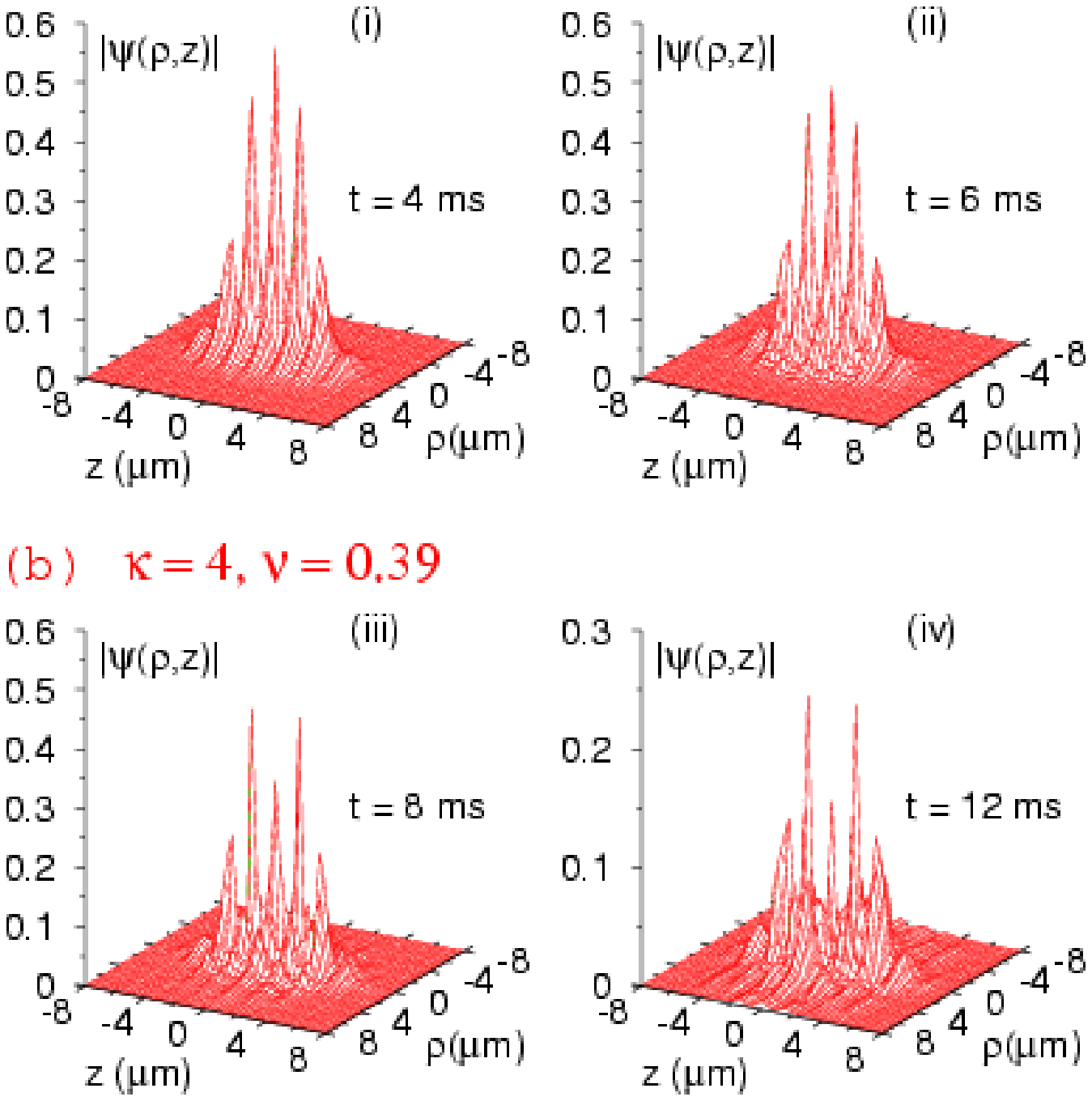}
\end{center}

\caption{A view of the evolution of the residual condensate wave function 
$|\psi(\rho,z)|$ in arbitrary units 
for initial scattering length  $a_{\mbox{in}}=
7a_0$, final scattering length   $a_{\mbox{col}}=
-30a_0$, initial number of atoms $N_0=16000$ at times $t=$   (i) 4 ms,
(ii) 6 ms, (iii) 8 ms and (iv) 12 ms for (a)  $\kappa =0$, $\nu=0.39$ and 
(b) $\kappa = 4$,  $\nu=0.39$.}  
\end{figure}

\begin{figure}%[!ht]
 
\begin{center}
\includegraphics[width=0.7\linewidth]{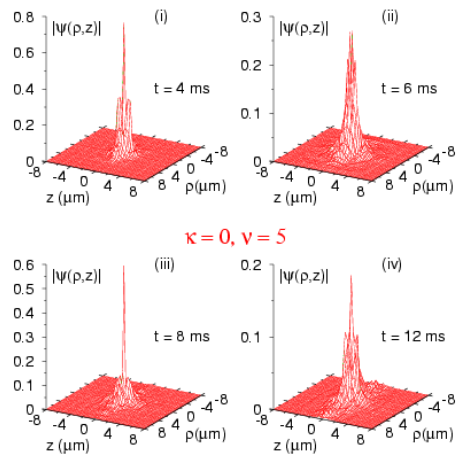}
\end{center}

\caption{Same as in Figs.  3 
for   $\kappa =0$, $\nu=5$.}

\end{figure}

Another aspect of Figs. 1 worth mentioning is that the number of atoms
in the remnant condensate  after the first explosion is larger in the
presence 
of an optical-lattice potential. Due to optical-lattice barriers one
essentially has local 
collapse of different pieces of the condensate in this
case
as opposed to a global collapse to the center of the condensate in the
case of a harmonic trap alone. Consequently, the collapse
is more violent with greater loss of atoms 
in the absence of an optical-lattice trap. This is why the
remnant number after the first explosion is larger in the presence of an
optical trap.

\subsection{Evolution of the Shape of the Condensate}

Next we consider the evolution of the shape of the residual condensate. In
Fig. 3 (a) we show the profile of the wave function $\psi(\rho,z)$ at
different times during explosion for $N_0=16000$, $\kappa =0$, $\nu=0.39$,
$a_{\mbox{in}}= 7a_0$, and $a_{\mbox{col}} = -30a_0$.  This is the case of
a cigar-shaped condensate used in the experiment \cite{don}. During
explosion the condensate wave function develops a three-peak structure
noted before in \cite{th3}. In Fig. 3 (b) we illustrate the profile of
the wave function $|\psi(\rho,z)|$ at different times during explosion of
the condensate formed in an optical-lattice potential with $\kappa =4$ in 
addition to the axial harmonic trap:
other parameters remaining the same as in Fig. 3 (a). The condensate now
develops a distinct multi-peak structure along the optical lattice in
place of the
three-peak structure in the absence of the optical-lattice
potential. However,
the number of peaks in the wave function is less by a factor of two to
three  than the number of
pits of the optical-lattice potential.  The number of distinct peaks in
the
wave
function in this case is five as can be seen in Fig. 3 (b).

The above distinct peaks in the wave function in the presence of the
optical-lattice potential may have interesting application in the
generation of radially bound and axially free bright solitons. The wave
function
of Fig. 3 (b) is axially bound. However, if the axial trap and the
optical-lattice potential are removed, or better an expulsive potential is
applied in the axial direction, the wave function will expand
axially.  The
side 
peaks of the wave function can evolve into  separate solitons and come
out in the axial direction which can be used as bright solitons in other
experiments.

The scenario of the evolution of the condensate is entirely different for
pancake-shaped condensate with $\nu > 1$. In that case  the condensate is
squezeed in the axial direction and a single  peak, rather than multiple
peaks, is formed in the
condensate wave function. This is illustrated in
Fig. 4 where we plot the condensate wave function for $\kappa =0$ and
$\nu =5$, the other parameters of simulation being the same as in Figs.
2. The use of optical-lattice potential in this case  also does not lead
to prominent peaks in the wave function in the axial direction.

\begin{figure}%[!ht]
 
\begin{center}
\includegraphics[width=0.7\linewidth]{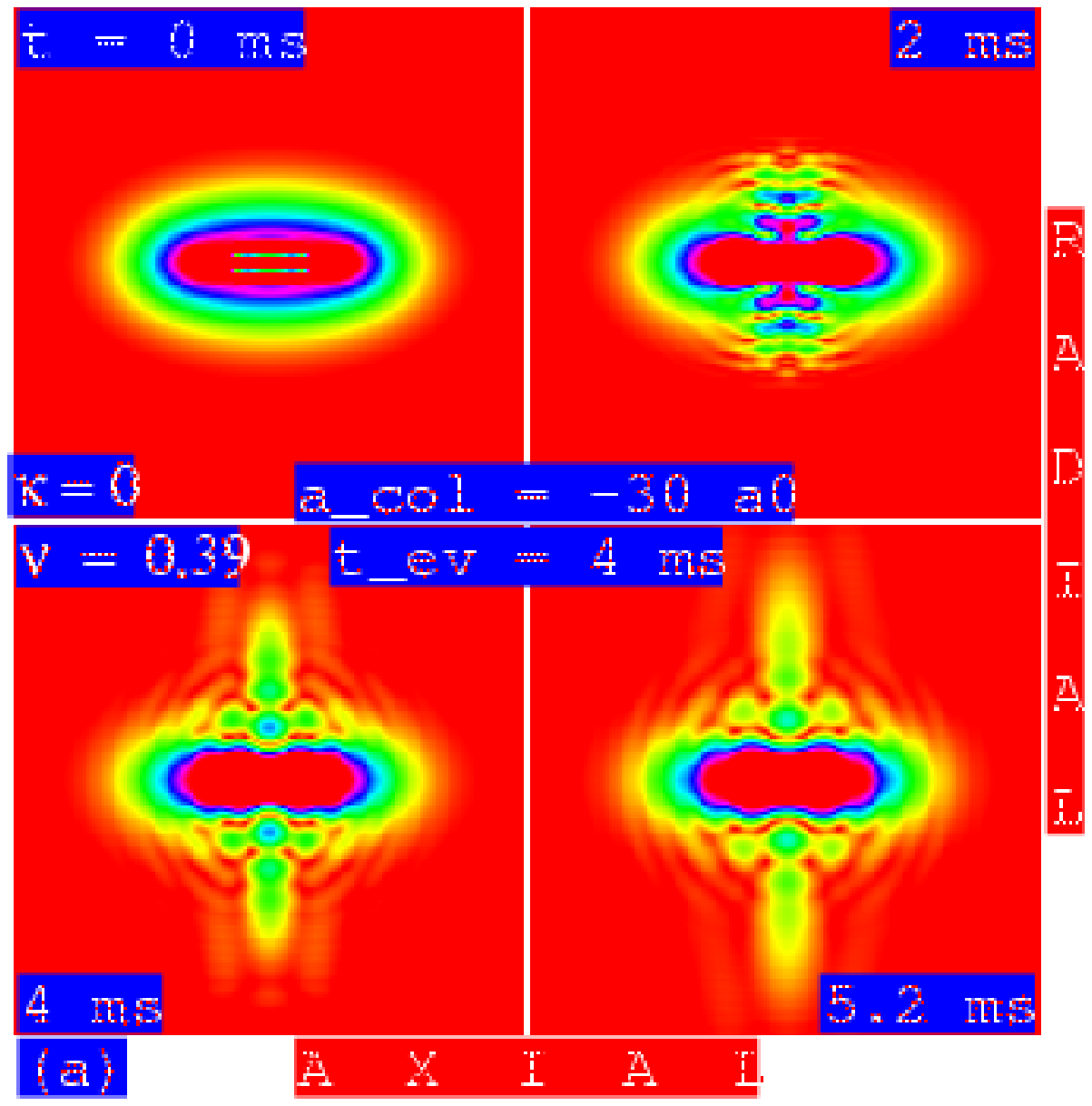}\hskip .2cm
\includegraphics[width=0.7\linewidth]{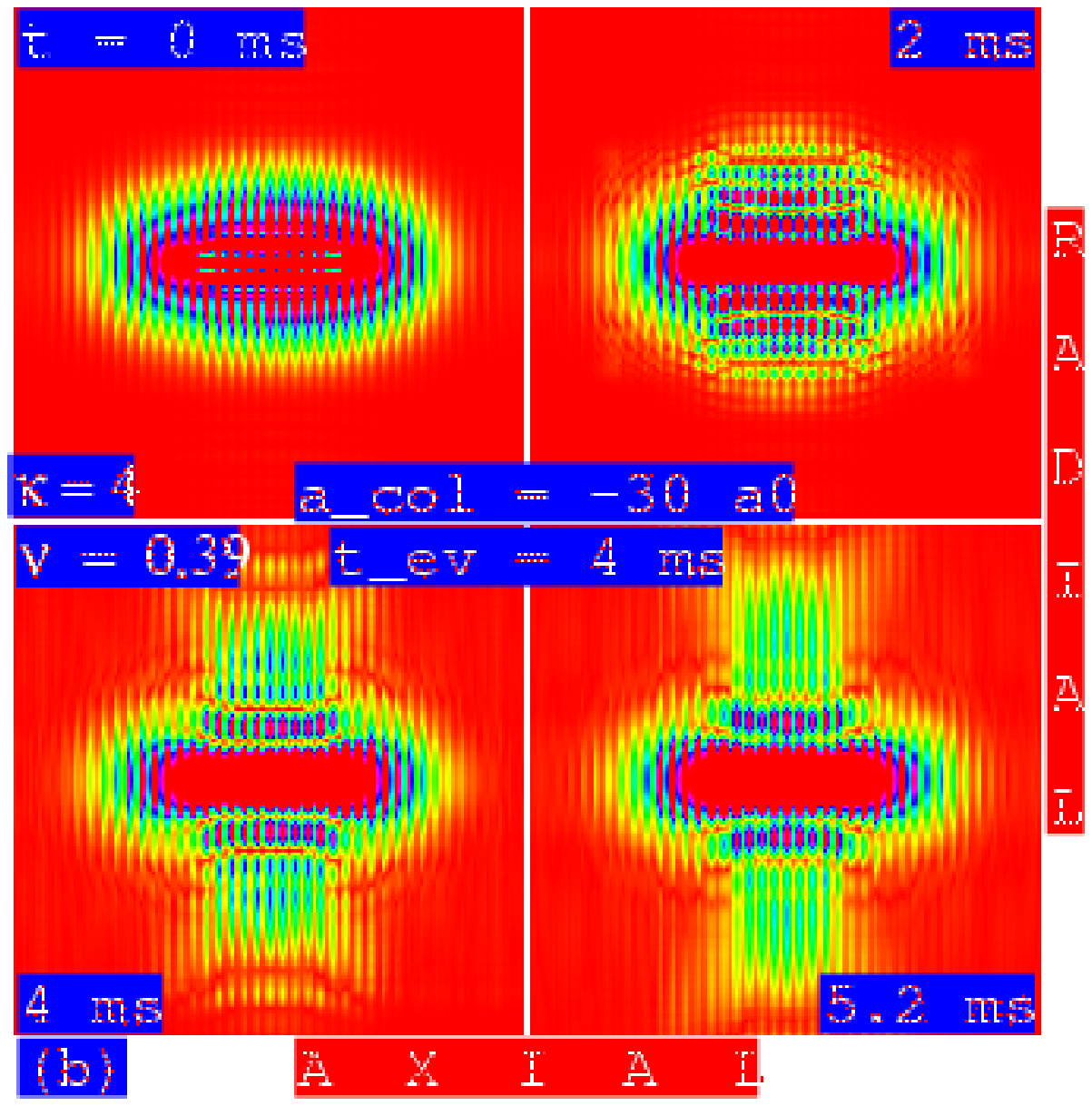}
\end{center}

\caption{(Color online)  A view of the evolution of radial jet at times
$t= 0,$ 2 ms, 4 ms
and 5.2 ms on a mat of size 16 $\mu$m $\times$  16 $\mu$m from a contour
plot of $|\psi(\rho,z)|$ for initial scattering length
$a_{\mbox{in}}=7a_0$, final  scattering length       $a_{\mbox{col}}=
-30a_0$, initial number of atoms $N_0=16000$,  
(a) without an optical-lattice potential ($\kappa=0$) and (b) 
with  an optical-lattice potential ($\kappa=4$). In both cases the jet
formation was started by jumping the scattering length to
$a_{\mbox{quench}}=0$ after a time $t_{\mbox{ev}}=4$ ms of the beginning
of
collapse.}
\end{figure}

\begin{figure}%[!ht]
 
\begin{center}
\includegraphics[width=0.7\linewidth]{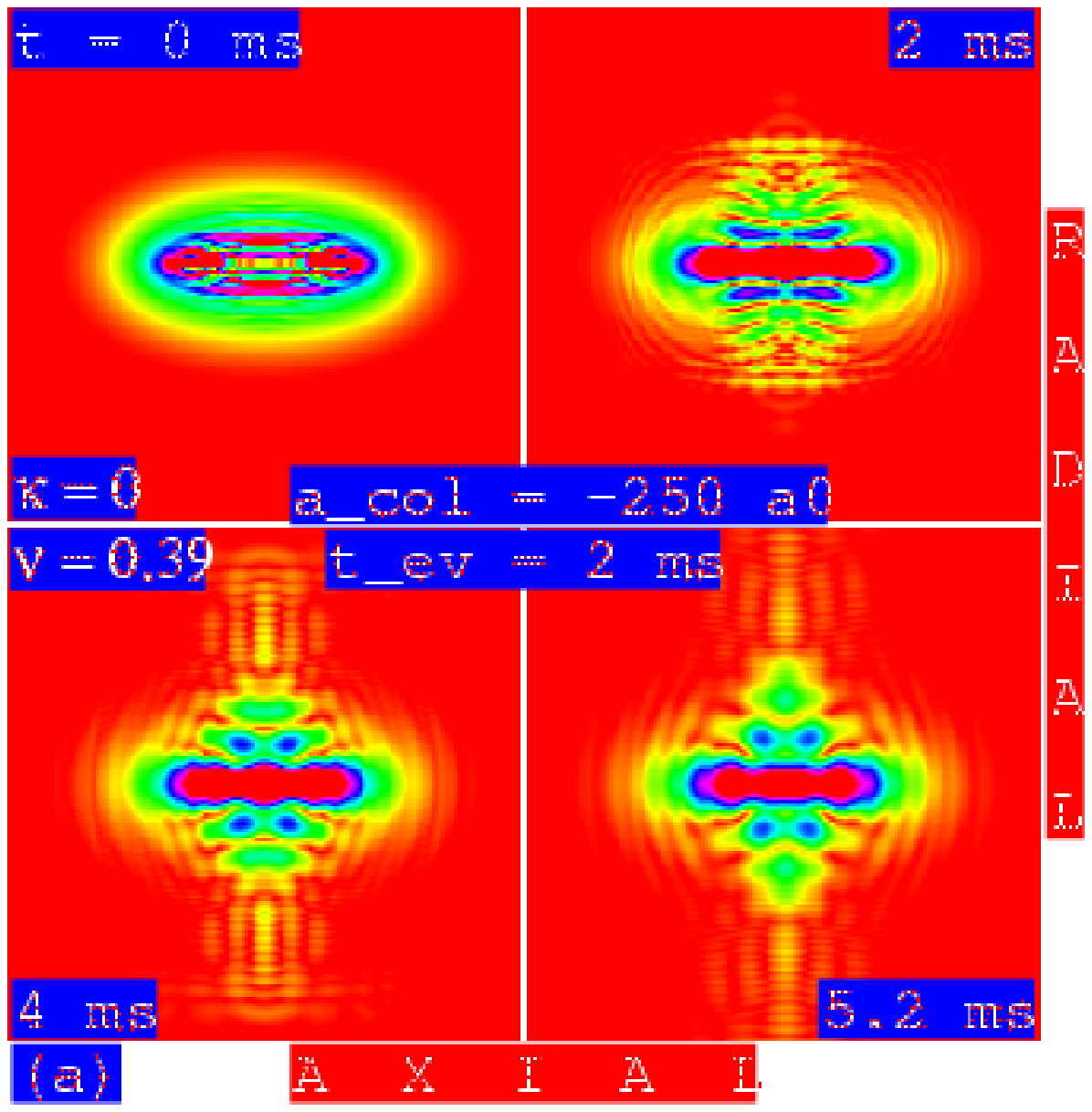}\hskip .2cm
\includegraphics[width=0.7\linewidth]{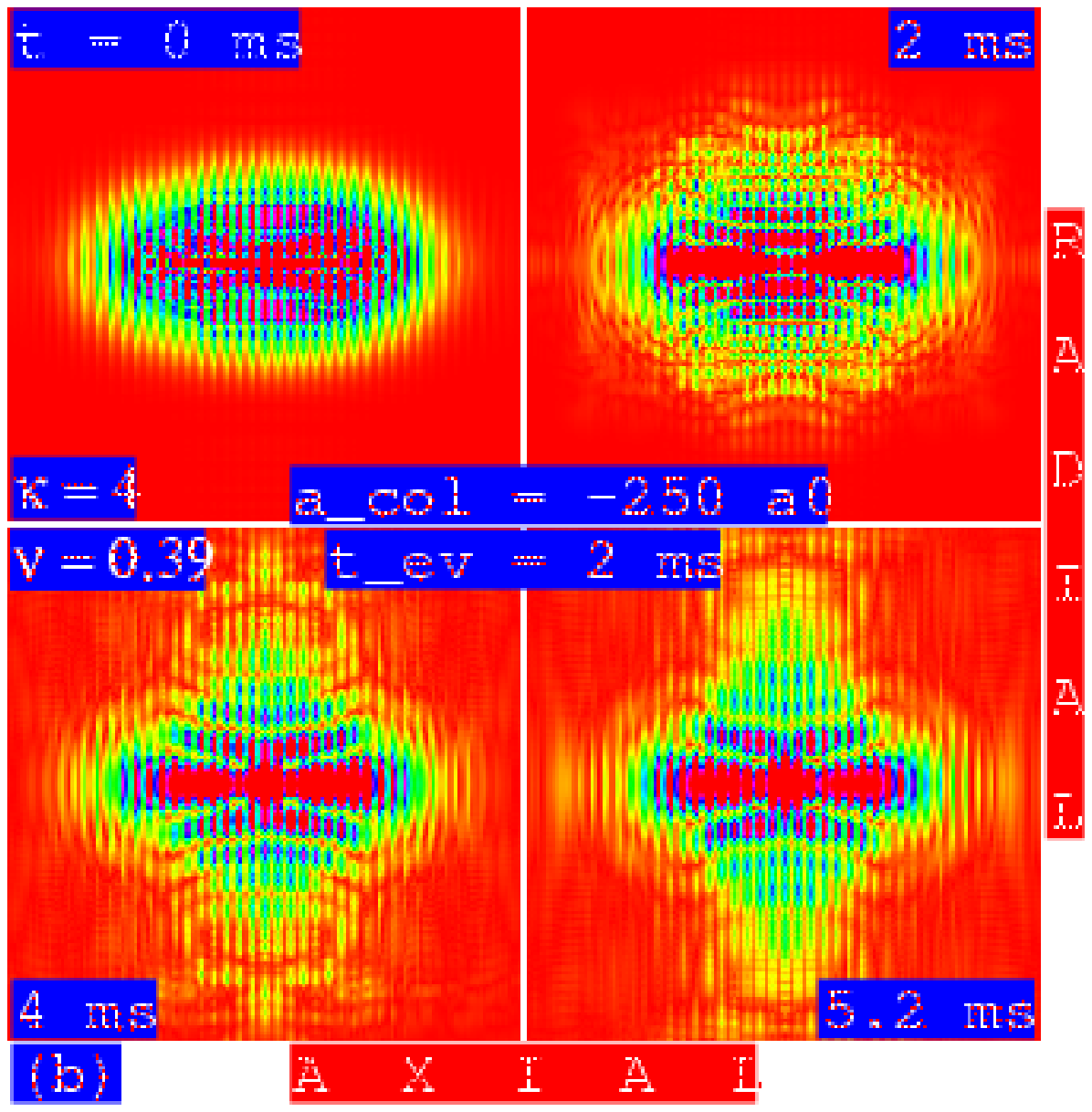}
\end{center}

\caption{(Color online)  Same as in Figs. 5 for $a_{\mbox{col}}=-250a_0$, 
and $t_{\mbox{ev}}=2$ ms.
}
\end{figure}

\subsection{Jet Formation}

Another interesting feature of the experiment of Donley {\it et
al} \cite{don}
is the formation of jet. 
As the collapse was 
suddenly terminated after an evolution time $t_{\mbox{ev}}$ by jumping the
scattering length from $a_{\mbox{col}}$ to $a_{\mbox{quench}} \ge 0$, the
jet atoms were slowly formed in the radial direction.
In the strongly collapsing condensate, 
local radial  spikes are formed during particle loss as can be
seen from a plot of the numerically calculated wave function
\cite{th2} and in experiment \cite{don}. 
During particle loss the top of the spikes 
are torn and ejected out and new spikes are formed until the
explosion and particle loss are over. There is a
balance between central atomic attractive force and the outward kinetic
pressure. If the attractive force is now suddenly removed by stopping the
collapse by applying $a_{\mbox{quench}}=0$, the highly collapsed 
condensate expands due to 
kinetic pressure,
becomes larger  and the recombination of atoms is greatly reduced. 
Consequently, the spikes expand and develop into a prominent jet
\cite{don}.

\begin{figure}%[!ht]
 
\begin{center}
\includegraphics[width=0.7\linewidth]{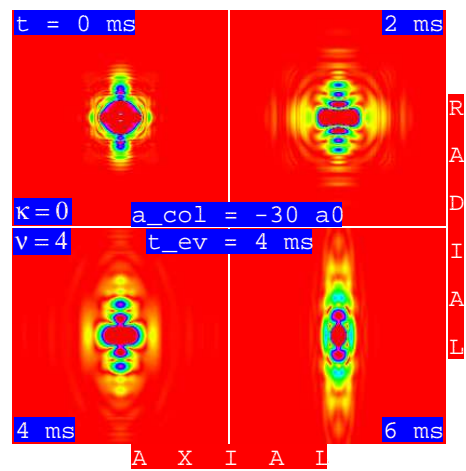}
%\postscript{fig1.eps}{1.}
\end{center}

\caption{(Color online)  Same as in Fig. 5 (a) for $\nu =4$
and $t=0$, 2 ms, 4 ms and 6 ms. 
}
\end{figure}

Now we
consider the jet formation as in the experiment of Donley {\it et al.}
\cite{don}
at  different
times $t$ of the collapsing condensate after jumping the 
scattering length suddenly from  $a_{\mbox{col}}=-30a_0$
to $a_{\mbox{quench}}= 0$ during explosion at time $t_{\mbox{ev}}$
from  the beginning of collapse and explosion.  The initial scattering
length
 $a_{\mbox{in}}=7a_0$ and number of atoms $N_0=16000$.  In Figs. 5 (a)
and (b)  we show the contour plot of the condensate for $t_{\mbox{ev}} =
2$ ms, without ($\kappa=0$)  and with ($\kappa=4$) an optical-lattice
potential, respectively, at different times $t=0$, 2 ms, 4 ms, and 5.2 ms
after jumping the scattering length to $a_{\mbox{quench}}= 0$.  A
prominent radial jet is formed slowly  at time $t=4 -6 $ ms after stopping
the
collapse at $t_{\mbox{ev}} = 4$ ms. 
The jet is much less pronounced for $t_{\mbox{ev}} = 2, 8$ ms
and 10 ms  (not shown in figure)
compared to the jet in Figs. 5.  There is a fundamental difference
between the jets in Figs. 5 (a) and (b) in the absence and presence of
optical-lattice potential. In Fig. 5 (a) the absence of the optical
potential the jet is narrow, whereas in Fig. 5 (b) it is wide and spread
over a number of optical lattice sites.

In addition, we studied jet formation for  different values 
of $a_{\mbox{col}}$ in place of $a_{\mbox{col}}=-30a_0$
and find that the general scenario remains similar. For example,  for
$a_{\mbox{col}}=-250a_0$, the collapse  and subsequent explosion
starts   at a small time. So for a good formation of jet a smaller value
of
$t_{\mbox{ev}}$ is to be preferred. In Figs. 6 we show the jet formation
for $a_{\mbox{col}}=-250a_0$ and $t_{\mbox{ev}}=2$ ms. In this case the
shape of the jet is different from that in Figs. 5. However, as in
Figs. 5, the jet gets broadened in the presence of the optical-lattice
potential.

Next we study the effect of the axial trap symmetry on jet formation. In
Figs. 5 and 6 the harmonic trap has cigar symmetry ($\nu =0.39 <1$). If
it is changed to pancake symmetry $(\nu >1)$, the condensate and the jet
gets compressed in the axial direction. Consequently, the radial jet 
is not very pronounced and can
not be clearly distinguished from the condensate. This is illustrated in
Fig. 7 for pancake-shaped trap with 
$\nu =4$    in the absence of an optical-lattice potential.  
In this case at $t= 6$ ms the condensate is more extended in the radial
direction compared to the condensate at $t=0$
due to the formation of jet. However, due to the overall compression of
the condensate in the axial direction the jet can not be easily separated
from the condensate. In contrast in Figs. 5 the jet is easily separated
from the central condensate. So pancake symmetry is not ideal 
for the study of a
jet. The situation does not change in the presence of an
optical-lattice potential superposed on a pancake-shaped trap.

\section{Conclusion}

In conclusion, we have employed a numerical simulation based on the
accurate solution \cite{sk1} of the mean-field Gross-Pitaevskii equation
with a cylindrical trap to study the evolution of a collapsing and
exploding condensate as in the experiment of Donley {\it et al.}
\cite{don}. In the GP equation we include a quintic three-body
nonlinear recombination loss term \cite{th1} that accounts for the decay
of the strongly attractive condensate.  
We also extend our investigation to different trap symmetries
and including an optical-lattice potential in the axial direction. In
addition
to studying the evolution of the size and the shape of the condensate, we
also study 
 the jet formation as observed  experimentally. Without any adjustable
parameter
the result of the present and previous simulations of this author 
are  in good agreement with some aspects of the  experiment by
Donley {\it et al.} \cite{don}. 

It is interesting to emphasize that the GP equation  does describe some
but not all aspects of the collapse experiment by Donley {\it et al.} 
and its
predictions for the ``time to collapse"  do not agree well with 
experiment \cite{th5,th11}. The failure to explain time to collapse is
dramatic as
intuitively one should expect the mean-field model to be a faithful model
for time to collapse involving the dynamics of the coldest atoms in the
condensate.  
However, there are aspects of experiments
which cannot be described by mean-field models \cite{th7,th9,th11,sk5},
e. g., the dynamics of missing and burst atoms \cite{don}. 
 Furthermore all numerical studies of this
experiment suffer from limited knowledge of the three-body loss
rate $K_{3}$ and even though many experimental features can be
described by a suitable choice of $K_{3}$, no value of  $K_{3}$ yields
simultaneous agreement between predictions of the GP equation and
all observable quantities of the experiment. In this situation 
it
would be of
great theoretical and experimental interests to see if
a
repeated experiment with different trap  parameters (and also including
an optical-lattice potential) 
would help to
understand the underlying physics and could make use of numerical
studies such as those described in the present paper.

\acknowledgments 

The work is supported in part by the CNPq 
of Brazil.

%\section*{References}

\end{document}